\documentclass[3p,times,procedia]{elsarticle}
\usepackage{nupha_ecrc}

\usepackage{amssymb}
\usepackage{amsmath, amssymb, graphics, setspace}
\usepackage{placeins}
\usepackage{tensor}
\usepackage{color}

\usepackage[normalem]{ulem}  

\renewcommand\sout{\bgroup \color{red} \ULdepth=-.5ex \ULset}

\volume{00}

\firstpage{1}

\journalname{Nuclear Physics A}

\runauth{}

\jid{nupha}

\jnltitlelogo{Nuclear Physics A}

\usepackage{amssymb}

\usepackage[figuresright]{rotating}

\begin{document}

\begin{frontmatter}



\dochead{XXVIIIth International Conference on Ultrarelativistic Nucleus-Nucleus Collisions\\ (Quark Matter 2019)}

\title{Local spin polarizations in relativistic heavy ion collisions}

\author[label1,label2]{Shuai Y. F. Liu}
\author[label3]{Yifeng Sun}
\author[label2]{Che Ming Ko}
\address[label1]{Quark Matter Research Center, Institute of Modern Physics, Chinese Academy of Science,Lanzhou, Gansu 730000}
\address[label2]{Cyclotron institute, Texas A\&M University, College Station, TX 77843-3366, USA}
\address[label3]{Laboratori Nazionali del Sud, INFN-LNS, Via S. Sofia 62, I-95123 Catania, Italy}




\begin{abstract}
Based on a generalized side-jump formalism for massless chiral fermions, which naturally takes into account the spin-orbit coupling in the scattering of two chiral fermions and the chiral vortical effect in a rotating chiral fermion matter, we have developed a covariant and total angular momentum conserved chiral transport model to study both the global and local polarizations of this matter.  For a system of massless quarks of random spin orientations and finite vorticity in a box, we have demonstrated that the model can exactly conserve the total angular momentum of the system and dynamically generate the quark spin polarization expected from a thermally equilibrated quark matter. Using this model to study the spin polarization in relativistic heavy-ion collision, we have found that the local quark spin polarizations depend strongly on the reference frame where they are evaluated as a result of the nontrivial axial charge distribution caused by the chiral vortical effect. We have further shown that because of the anomalous orbital or side-jump contribution to the quark spin polarization, the local quark polarizations calculated in the medium rest frame are qualitatively consistent with the local polarizations of Lambda hyperons measured in experiments.
\end{abstract}


\end{frontmatter}


\section{Introduction}\label{sec_intro}

The study of Lambda polarization in heavy-ion collision experiments at RHIC~\cite{STAR:2017ckg} opens a new window to study the properties of quark-gluon plasma. While the measured global Lambda polarization is well-understood in studies based on the  statistical approach~\cite{Becattini:2013fla,Pang:2016igs}, the measured Lambda local polarizations~\cite{Niida:2018hfw,Adam:2019srw} disagree with the predictions from these studies~\cite{Becattini:2017gcx,Xia:2018tes}, leading to the so-called the spin puzzle in relativistic heavy-ion collisions. Within the frameworks of the hydrodynamic and the statistical model, there have been many attempts to  solve this puzzle without much success.  These include ambiguities in the definition of the local spin polarization density in the thermal limit~\cite{Becattini:2018duy}, the introduction of spins in hydrodynamics~\cite{Florkowski:2019qdp,Bhadury:2020puc}, feed-down effects~\cite{Xia:2019fjf}, and different choices of the vorticity field~\cite{Florkowski:2019voj,Wu:2019eyi}. In studies base on the chiral kinetic approach~\cite{Sun:2017xhx,Sun:2018bjl}, both predicted global spin polarization and sign of the longitudinal local polarization ${\cal P}_z$ along the beam direction are consistent with the experimental observations, but  they fail to describe the measured local polarization ${\cal P}_y$ along the global angular momentum direction. Although the chiral kinetic approach in Refs.~\cite{Sun:2017xhx,Sun:2018bjl} has given very encouraging results, it lacks the conservation of total angular momentum, is non-covariant, and treats the chiral vortical effect ambiguously. These shortcomings have recently been overcome in our work~\cite{Liu:2019krs} through the development of an angular moment conserved covariant chiral kinetic approach based on a generalized side-jump formalism that can naturally take into account the spin-orbit coupling in the scattering of two massless quarks and the chiral vortical effect in a rotating quark matter. In the present proceedings, we briefly review this new approach and show that it gives a qualitatively correct description of the local polarizations ${\cal P}_y$ and ${\cal P}_z$ of Lambda hyperons measured in experiments.

\section{Formalism}

The  chiral kinetic theory with the side-jump effect is well-known for scattering of chiral fermions at zero impact parameter~\cite{Chen:2014cla,Chen:2015gta} and can be  generalized to the case of scattering at finite impact parameter as shown in Ref.~\cite{Liu:2019krs}, which is needed for the scattering of massless quarks in relativistic heavy ion collisions. In this approach, the covariant angular momentum tensor $J^{\mu\nu}$ of a chiral fermion is given by the sum of an orbital part $L^{\mu\nu}=x^{\mu}p^{\nu}-x^{\nu}p^{\mu}$ and a spin part
$S^{\mu\nu}=\lambda{\frac{\epsilon^{\mu\nu\alpha\beta} p_\alpha n_\beta}{p\cdot n}}$, where the $x^\mu$ and $ p^\mu $ are its four  coordinate and momentum, respectively, $\lambda=\pm 1/2$ is its helicity, and $n=(1,\textbf{0})$ to ensure the alignment of its spin and momentum directions. Although the total angular momentum and momentum of the chiral particle transform under a Lorentz transformation $\Lambda$ as usual, i.e., $J^\prime=\Lambda J\Lambda^T$ and $p^\prime=\Lambda p$, to maintain its helicity requires, however, an additional side jump $\Delta$ in its coordinate, resulting in the transformation
\begin{align}
x'^{\mu}=\Lambda\indices{^\mu_\alpha} x^\alpha+\Delta^{\mu}_{{\tilde n}n'},\quad \Delta^{\mu}_{{\tilde n}n'}=\lambda \frac{\epsilon^{\mu\alpha\beta\gamma}p'_{\alpha}{\tilde n}_{\beta}n_{\gamma}'}{(p'\cdot {\tilde n})(p'\cdot n')},
\label{eq_jumptrans}
\end{align}
with ${\tilde n}^\mu=\Lambda\indices{^\mu_\alpha}(1,\textbf{0})_\alpha =\Lambda\indices{^\mu_0}$ and $n^\prime=(1,{\bf 0})$.

For two chiral fermions with initial momenta $ \textbf{p}_1=-\textbf{p}_2=\textbf{p} $ and positions $ \textbf{x}_1=\textbf{X}+\mathbf{x}/2$ and $ \textbf{x}_2=\textbf{X}-\mathbf{x}/2$ in their center of mass (CM) frame, their momenta $ \textbf{p}_1^\prime=-\textbf{p}_2^\prime=\textbf{p}^\prime $ and positions $ \textbf{x}^\prime _1=\textbf{X}^\prime+\mathbf{x^\prime/2}$ and $ \textbf{x}_2^\prime=\textbf{X}^\prime-\mathbf{x^\prime}/2$ after the scattering can be determined from the conservation of  the spatial components $J^{ij}$ and time components $J^{i0}$ of their total angular momentum tensor, i.e., $\textbf{p}'=R_{\hat{\textbf{J}}}(\phi)\textbf{p}$,  
$\mathbf{x}'=R_{\hat{\textbf{J}}}(\phi)\mathbf{x}$, and $\textbf{X}^\prime=\textbf{X}-(\textbf{p}-\textbf{p}^\prime)dt/\sqrt{s}$, where $ R_{\hat{\textbf{J}}}(\phi) $ is the rotation operator around the direction of total orbital angular momentum $ \textbf{J}=\sum_{i=1,2}(\textbf{x}_i\times \textbf{p}_i+\lambda \hat{\textbf{p}}_i) $ by an angle $\phi$. The factor $dt$ in the last expression is due to the time difference between the two chiral fermions in their CM frame if we take their scattering times to be same in the laboratory (LAB) frame. The positions and momenta of the two chiral chiral fermions in the LAB frame after the scattering can then be obtained using  $ \tilde{n}=(p_1+p_2)/\sqrt{s}$ in Eq.~(\ref{eq_jumptrans}) to ensure the conservation of their total angular momentum in the LAB frame as well. 

Based on the above described scattering between two chiral fermions, we have constructed in Ref.~\cite{Liu:2019krs} an angular momentum conserved covariant transport model for massless quarks by treating the quark-quark scattering cross section as a parameter and allowing them to freely propagate  between scatterings. With the phase-space distribution functions $ f_\lambda(x,p)$ of left- and right-handed quarks obtained from this transport model, the average quark polarization $\mathcal{P}$ can be calculated from the quark axial current density $J_5^\mu$ and the time component of the current density $n(x)=j^0(x)$ as\begin{align}
\mathcal{P}=\int d^3x\textbf{j}_5(x)/\int  d^3x \,n(x), \quad
j_5^\mu=\sum_{\lambda=\pm 1/2}\int \frac{d^3\textbf{p}}{(2\pi)^3p}\lambda p^\mu f_\lambda+\lambda S^{\mu\nu}\partial_\nu f_\lambda. 
\label{eq_j5_pol}
\end{align}
In the above, the first term $ \lambda p^\mu f_\lambda $ is the usual spin polarization and is denoted as the spin term in Ref.~\cite{Liu:2019krs}, and the second term $ \lambda S^{\mu\nu}\partial_\nu f_\lambda $ is due to the side-jump effect required by the covariance of the axial current. Since the latter is proportional to the local vorticity in the quark matter and thus related to the orbital motions of quarks, it is called the orbital term in Ref.~\cite{Liu:2019krs}. With the quark polarization $\mathcal{P}$  calculated in terms of the spatial and time components of the covariant axial current $j_5^\mu$ and vector current$ j^\mu $, its value in any frame can be obtained from the corresponding Lorentz transformation of the covariant currents.


\section{Results}

\begin{figure*}[tbh]
	\centering
	\includegraphics[width=0.6\columnwidth]{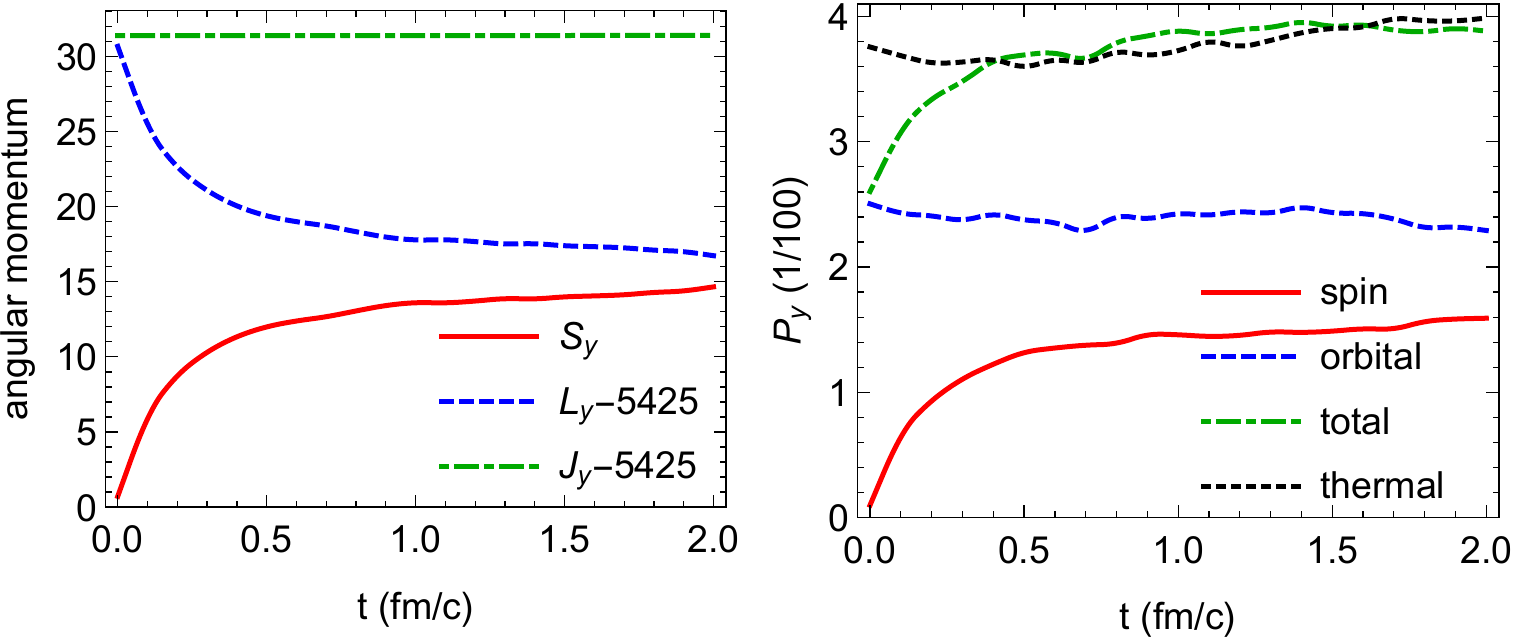}
	\vspace{-0.3cm}
	\caption{Left window: the spin $S_y$, orbital $L_y$, and total $J_y$ angular momenta in the $y$-direction. Right window: total polarization in the $y$-direction and its spin and orbital contributions together with the polarization expected from the thermal model.}
	\label{fig_bench}
\end{figure*}

Using the angular momentum conserved covariant chiral transport model with an isotropic quark-quark scattering cross section of 10 mb, we have studied the time evolution of a box of initially unpolarized quark matter rotating along the $y$-axis or having a finite vorticity in this direction as it expands.  As shown in the left window of Fig.~\ref{fig_bench}, the total angular momentum (dash-dotted line) in the system is conserved exactly during its expansion, and its spin part (solid line) increases at the expense of the decrease of its orbital part (dash line). For the time evolution of the polarization in the quark matter, shown in the right window, both the spin (solid line) and orbital (dash line) contributions are important, and the inclusion of both is essential to reproduce the spin polarization $\omega/(2T)$ expected from the thermal model (short-dash line), where $\omega$ and $T$ are the angular velocity or vorticity and temperature of the quark matter, respectively.

\begin{figure*} [tbh]
	\centering
	\includegraphics[width=0.59\columnwidth]{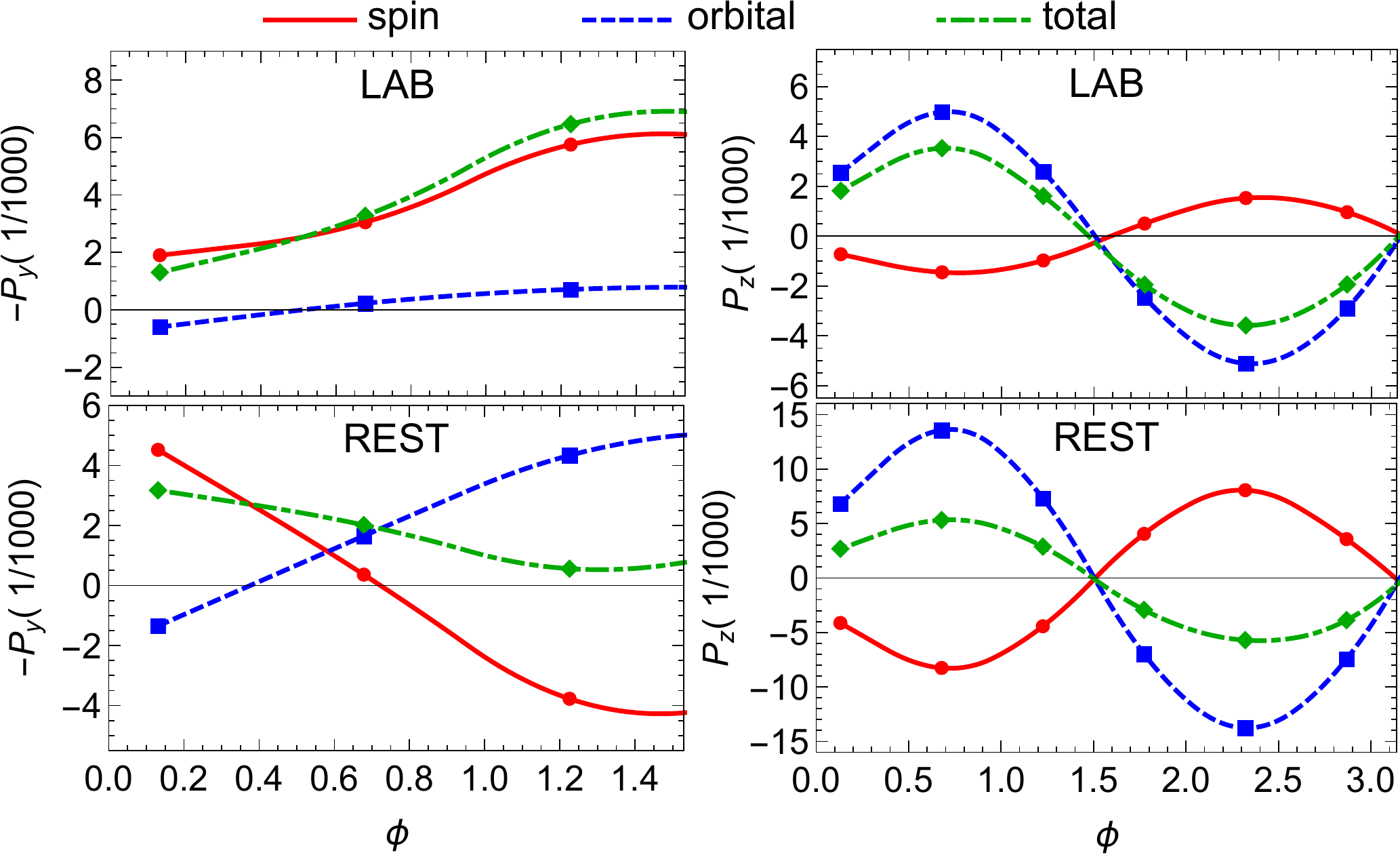}
	\includegraphics[width=0.4\columnwidth]{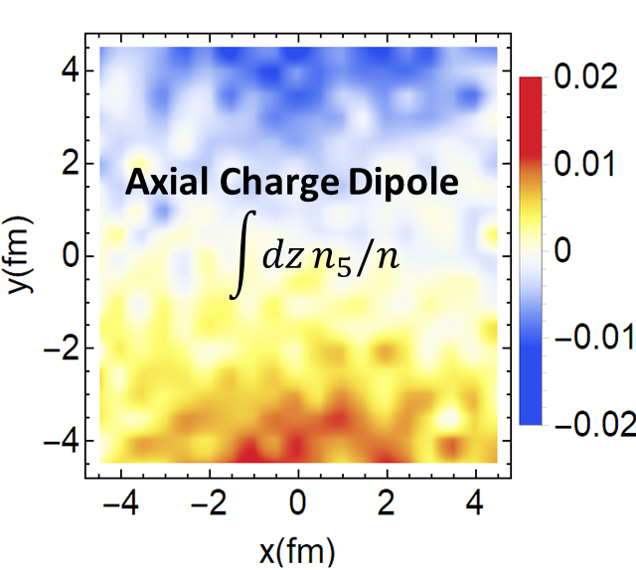}
	\caption{Left and middle windows: local quark polarizations ${\cal P}_y$ perpendicular to the reaction plane and along the beam direction ${\cal P}_z$ as a function of the azimuthal angle $\phi$ in the transverse plane calculated in the fireball (LAB) frame and the medium rest (REST) frame. Right window: quark axial charge distribution in the transverse plane.}
	\label{fig_AuAu}
\end{figure*}

We have also used the angular momentum conserved covariant chiral transport model to study the polarization of the quarks produced in non-central Au-Au $\sqrt{s}=200~\text{GeV}$ at 30-50\% centrality with the initial quark phase-space distributions taken from the AMPT model~\cite{Lin:2004en}. The results after evolving the system for 8~fm/c are shown in  Fig.~\ref{fig_AuAu}  for the azimuthal angle dependence of the local polarizations ${\cal P}_y$ (left window) and ${\cal P}_z$ (middle window) in both the LAB or fireball frame and the medium rest (REST) frame. It is seen that both the magnitude and sign of the local polarizations depend on the reference frame where they are evaluated, especially for the spin contribution to the local polarizations. This frame dependence of quark local polarizations is due to the non-zero local axial charge density distributions in the transverse plane of the a collision, shown in the right window of Fig.~\ref{fig_AuAu}, which is caused by the chiral vortical effect in the quark matter as first pointed out in Ref.~\cite{Sun:2018bjl} based on a simple chiral transport approach.  Since the time and spatial components  of the axial current $ j_5 $, which are related to the axial charge density and the quark polarization, respectively, a Lorentz boost  can thus affect the quark local polarizations, particularly in an expanding quark matter.  As pointed out in Ref.~\cite{Liu:2019krs}, our results in the medium rest frame shown in Fig.~\ref{fig_AuAu}, where the $ \mathcal{P}_y $ is dominated by the spin part and the $\mathcal{P}_z $ is dominated by the orbital part, qualitatively agree with the experimentally measured local polarizations of Lambda hyperons ~\cite{Niida:2018hfw,Adam:2019srw}.

\section{Conclusion}

In summary, based on the side-jump formalism for the scattering of chiral fermions, we have developed a total angular momentum conserved covariant transport model for massless quarks. This model has allowed us to automatically include the spin-orbit coupling in the scattering of two massless quarks and the chiral vortical effect in the rotating quark matter produced in relativistic heavy ion collisions. Applying this model to a box of quark matter with nonzero vorticity, we have demonstrated that the total angular momentum of the system is rigorously conserved during its expansion and the final quark spin polarization, which includes contributions from both the spin of the quark and its orbital motion, agrees with the expected thermal limit. For non-central relativistic heavy-ion collisions, we have found that the quark local polarizations depend on the frame where they are evaluated as a result of the nontrivial axial charge distribution generated by the chiral vortical effect in the transverse plane of a heavy ion collision. With the spin part of the polarization dominating the quark local polarization perpendicular to the reaction plane and the orbital part dominating the quark local polarization along the beam direction, we have found that their dependence on the azimuthal angle in the transverse plane of a heavy ion collision is similar to those of Lambda hyperons measured in experiments if they are evaluated in the medium rest frame of the produced quark matter. Our study thus offers a plausible explanation for the unexpected behaviors of local Lambda spin polarizations measured in relativistic heavy ion collisions, compared to those predicted from the thermal model.  However, our study is based on the chiral limit of massless quarks. Since quarks have nonzero current masses, particularly for the strange quark that in the constituent quark model determines the spin of a Lambda hyperon, it is of great interest to extend our approach to quarks of finite masses and to also construct an angular-momentum conserved hadronization model to convert the quark polarizations to those of hadrons. Such a realistic model is expected to provide a more definitive explanation for the spin puzzle in relativistic heavy ion collisions.


\section{acknowledgments}

This work was supported in part by the US Department of Energy under Contract No. DE-SC0015266 and the Welch Foundation under Grant No. A-1358.



\bibliography{refcnew}
\bibliographystyle{h-elsevier}






\end{document}